# TestTDO v1.3's Terms, Properties, Relationships and Axioms - A Top-Domain Software Testing Ontology

**Guido Tebes, Denis Peppino, Pablo Becker** and **Luis Olsina**

GIDIS_Web, Facultad de Ingeniería, UNLPam, General Pico, LP, Argentina

[guido_tebes, olsinal, beckerp]@ing.unlpam.edu.ar; denispeppino92@gmail.com

**Abstract.** The present preprint specifies and defines all Terms, Properties, Relationships and Axioms of TestTDO (software Testing Top-Domain Ontology) v1.3, which is a slightly updated version of its predecessor, TestTDO v1.2. TestTDO is a top-domain ontology built in the context of a four-layered ontological architecture named FCD-OntoArch (Foundational, Core, and Domain Ontological Architecture for Sciences). This is a four-layered ontological architecture, which considers Foundational, Core, Domain and Instance levels. In turn, the domain level is split down in two sub-levels, namely: Top-domain and Low-domain ontological levels. Ontologies at the same level can be related to each other, except for the foundational level where only the ThingFO ontology is found. In addition, ontologies' terms and relationships at lower levels can be semantically enriched by ontologies' terms properties and relationships from the higher levels. Some TestTDO's terms and are extended primarily from SituationCO (Situation Core Ontology), and ProcessCO (Process Core Ontology) concepts. Stereotypes are the mechanism used for enriching TestTDO terms. Note that annotations of updates from the previous version (v1.2) to the current one (v1.3) can be found in Appendix A.

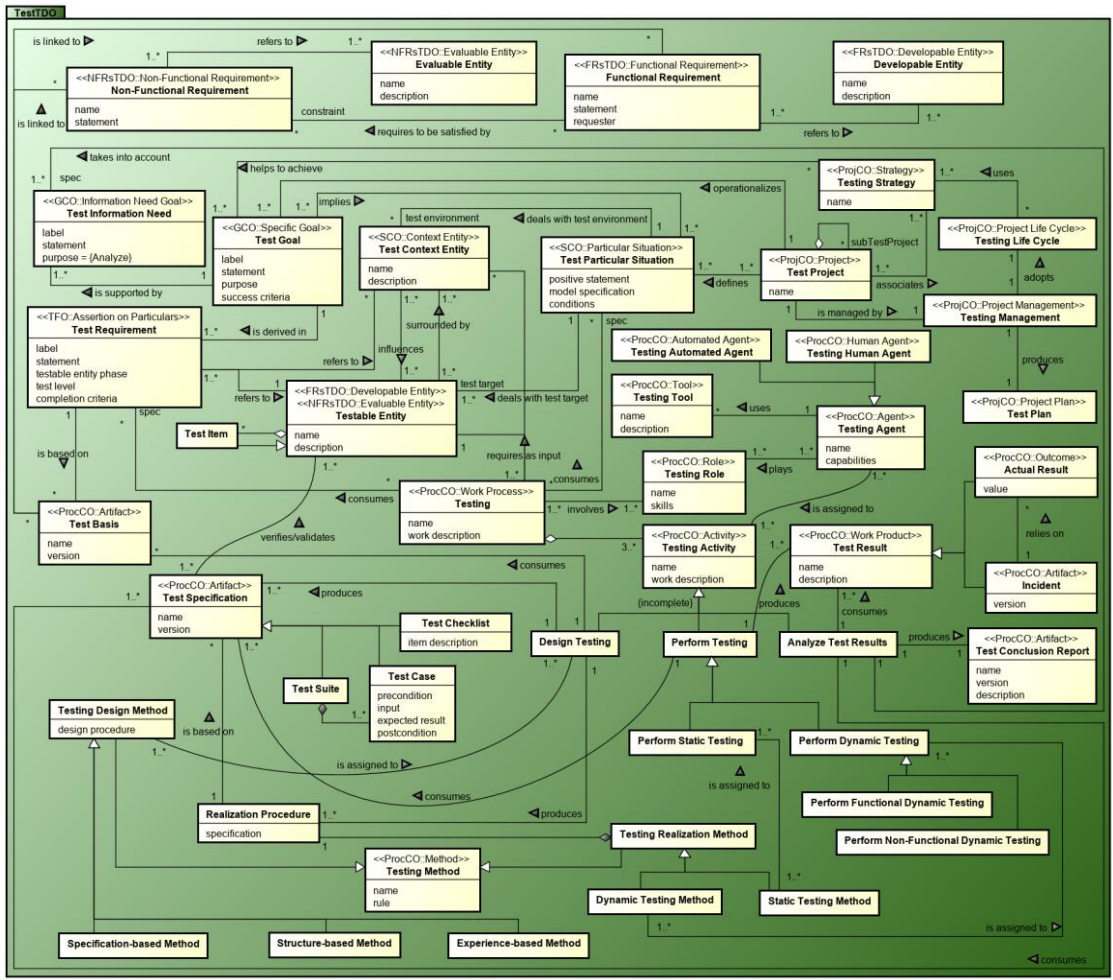

**Figure 1.** TestTDO v1.3: Top-Domain Ontology for Testing, which is placed at the domain level of FCD-OntoArch [1] in Figure 2. This is a revised version of TestTDO v1.2 [2], where annotations of changes/updates from the previous version (v1.2) to the current one (v1.3) can be found in Appendix A. Note that TFO stands for Thing Foundational Ontology [3], SCO for Situation Core Ontology [3], ProcCO for Process Core Ontology [4], ProjCO for Project Core Ontology [5], GCO for Goal Core Ontology [5], NFRsTDO for Non-Functional Requirements Top-Domain Ontology [6], [7], and FRsTDO for Functional Requirements TDO [7], [8].

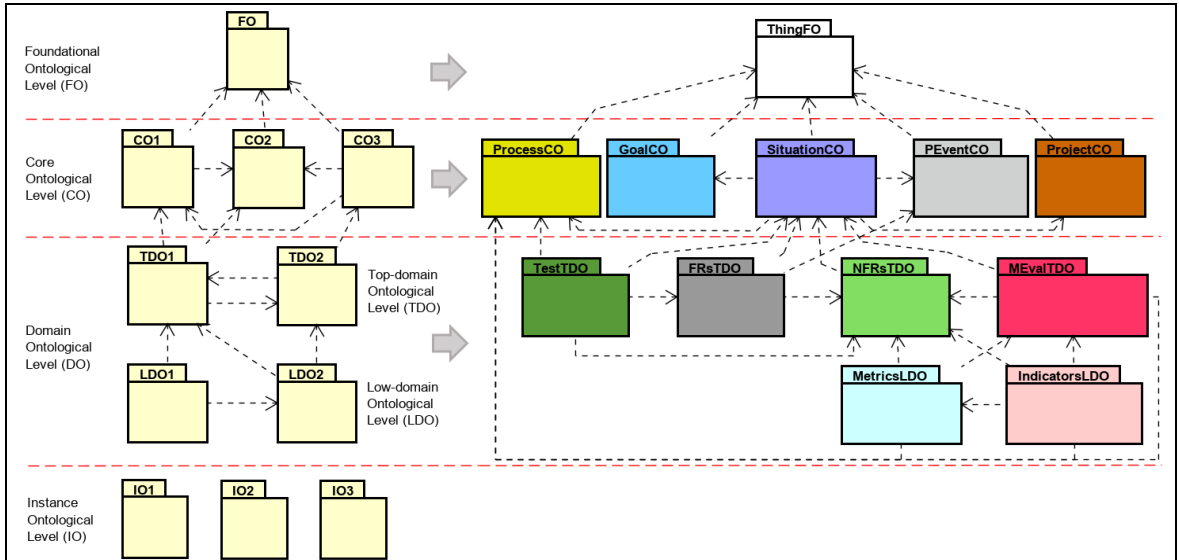

**Figure 2.** Allocating the TestTDO component or module in the context of the four-layered ontological architecture so-called FCD-OntoArch (*Foundational, Core, and Domain Ontological Architecture for Sciences*) [1].

## Testing Component – TestTDO v1.3's Terms

| Term | Definition |
|---|---|
| **Actual Result** | It is a Test Result that represents a numerical or categorical value (expected or unexpected). |
| **Analyze Test Results** <br> (synonym: Testing Analysis) | It is a Testing Activity that takes into account the specific Test Information Need in order to produce a Test Conclusion Report by consuming one or more Test Results and Test Specifications. |
| **Design Testing** <br> (synonym: Testing Design) | It is a Testing Activity aimed at designing a set of Test Specifications (i.e., Test Cases, Test Suites and/or Test Checklists) as well as Realization Procedures. |
| **Dynamic Testing Method** | It is a Testing Realization Method for a task included in a Perform Dynamic Testing activity. |
| **Experience-based Method** | It is a Testing Design Method that uses the Testing Human Agent's knowledge, expertise and intuition while enacting the Design Testing activity for deriving Test Specifications. <br><br> <u>Note</u>: Examples of Experience-based Methods and practices (as per [9]) are Error Guessing, Exploratory Testing, Software Attacks and Ad Hoc Testing, under the name of "*experience-based test design techniques*". |
| **Incident** <br> (synonym: Anomaly, Defect, Issue Report) | It is a Test Result that reports deviations (e.g. between the expected result and the Actual Result), anomalies (e.g. an error or a failure) or other arisen issues during the Perform Testing activity. |
| **Perform Dynamic Testing**<br><br>(synonym: Dynamic Testing) | It is a Perform Testing activity aimed at verifying/validating a Testable Entity against one or more Test Specifications with the execution of its software code.<br><br>Note 1: Dynamic Testing as per ISTQB v. 3.2 [10] is "*testing that involves the execution of the software of a component or system*".<br><br>Note 2: Dynamic Testing as per ISO/IEC/IEEE 29119-1 [9] is "*testing that requires the execution of the test item*". |
| **Perform Functional Dynamic Testing**<br><br>(synonym: Functional Dynamic Testing) | It is a Dynamic Testing activity aimed at verifying/validating the compliance of a Testable Entity with Functional Requirements [8].<br><br>Note: Functional Testing as per [10] is "*testing conducted to evaluate the compliance of a component or system with functional requirements*". |
| **Perform Non-functional Dynamic Testing**<br><br>(synonym: Non-functional Dynamic Testing) | It is a Dynamic Testing activity aimed at appraising the compliance of a Testable Entity with Non-Functional Requirements [6].<br><br>Note: Non-Functional Testing as per [10] is "*testing conducted to evaluate the compliance of a component or system with non-functional requirements*". |
| **Perform Static Testing**<br><br>(synonym: Static Testing) | It is a Perform Testing activity aimed at checking a Testable Entity against one or more Test Specifications without the execution of its software code (if any).<br><br>Note 1: Static Testing as per [10] is "*testing a work product without code being executed*".<br><br>Note 2: Static Testing as per [9] is "*testing in which a test item is examined against a set of quality or other criteria without code being executed*". |
| **Perform Testing**<br><br>(synonym: Testing Realization) | It is a Testing Activity aimed at enacting a Static or Dynamic Testing.<br><br>Note: This activity consumes one or more Test Specifications and produces one or more Test Results. |
| **Realization Procedure** | Arranged set of Testing Realization Method's instructions or operations, which specifies how must be performed the Perform Testing activity using the Test Specification.<br><br>Note: Mainly when Dynamic Testing that involves Test Cases execution is carried out, this term (i.e., Realization Procedure) can be synonymous with Test Procedure. A Test Procedure as per ISO/IEC/IEEE |
| | 29119-1 [9] is a "*sequence of test cases in execution order, and any associated actions that may be required to set up the initial preconditions and any wrap up activities post execution. Test procedures include detailed instructions for how to run a set of one or more test cases selected to be run consecutively, including set up of common preconditions, and providing input and evaluating the actual result for each included test case.*". |
| **Specification-based Method**<br><br>(synonym: Black-box Method) | It is a Testing Design Method that always uses a Test Basis while enacting the Design Testing activity for deriving Test Specifications without referring to the internal structure of the Testable Entity.<br><br><u>Note</u>: Examples of Specification-based Methods are Boundary Value Analysis, State Transition Testing, and Decision Table Testing, among others (covered in [11] under the name of "*specification-based test design techniques*"). |
| **Static Testing Method** | It is a Testing Realization Method for a task included in a Perform Static Testing activity.<br><br><u>Note</u>: Static Testing Method examples are Walkthrough, Technical Review, Inspection, etc. |
| **Structure-based Method**<br><br>(synonym: White-box Method) | It is a Testing Design Method that uses the internal structure of the Testable Entity, and sometimes also uses a Test Basis, while enacting the Design Testing activity for deriving Test Specifications.<br><br><u>Note</u>: Examples of Structure-based Methods are Branch Testing, Condition Testing and Data Flow Testing, among others (covered in [11] under the name of "*structure-based test design techniques*"). |
| **Test Basis** | Artifact consumed by a Design Testing activity for designing the Test Cases and Checklists.<br><br><u>Note</u>: Test Basis represents Artifacts [4] that may come from development and/or maintenance such as requirements specification, architectural design, documented source code, etc., which in turn could be linked to NFRs and FRs. |
| **Test Case** | It is a Test Specification that contains the necessary information (e.g. preconditions, inputs, expected results and postconditions) to perform mainly Dynamic Testing.<br><br><u>Note</u>: A Test Case, as per [10] is "*a set of preconditions, inputs, actions (where applicable), expected results and postconditions, developed based* |
| | *on test conditions*". |
| **Test Checklist** | It is a Test Specification that contains a list of items to be checked in order to perform mainly Static Testing. |
| **Test Conclusion Report** | It is an Artifact that documents the analysis of all Test Results. |
| **Test Context Entity** | It represents the concrete Context Object in which the Testable Entity is situated.<br><br>Note 1: Test Context Entity has a semantic of Context Entity [3].<br><br>Note 2: A Test Context Entity belongs to a Context Category, which test environment is one type of it.<br><br>Note 3: A test environment is "*an environment containing hardware, instrumentation, simulators, software tools, and other support elements needed to conduct a test*" as per [10]. |
| **Test Goal**<br><br>(synonym: Test Objective) | It is a business goal for testing that the organization intends to achieve.<br><br>Note: A business goal is, in turn, a goal, which is a "*The statement of the aim to be achieved by the organization, which considers the propositional content of a purpose in a given time frame and context*" [5]. |
| **Test Information Need** | It is an Information Need Goal [5] that is achieved by conducting Testing Activities.<br><br>Note: Particularly, a specific Test Information Need is taken into account by the Analyze Test Results Activity. |
| **Test Item** | A part of a Testable Entity, which is a Test Object as well. |
| **Test Particular Situation** | It represents an association between one or more Testable Entities in the role of test target and none or many Test Context Entities in the role of test environment.<br><br>Note: The term Test Particular Situation has semantic of Particular Situation [3], which is an Assertion on Particulars from ThingFO [3]. |
| **Test Plan** | The document (Artifact) that describes how the Test Project will be planned, scheduled, executed, monitored, and controlled. |
| **Test Project** | It is a Project [5] representing a temporary and goal-oriented endeavor for testing with definite start and finish dates, which considers a managed set of |
| | interrelated Testing Activities, tasks and resources aimed at producing and modifying unique work products (e.g., Test Specifications, Test Results, etc.) for satisfying a given requester need. |
| **Test Requirement** | It states, taking into account the Test Goal purpose, what must be verified/validated of a Testable Entity (and/or Test Item) based on the Test Basis, if any. |
| | Note 1: A Test Requirement must include the test level (e.g., unit, integration, system, acceptance, etc.) and the phase (e.g., development, operative, maintenance, etc.) of the Testable Entity. |
| | Note 2: The term Test Requirement has semantic of Assertion on Particulars from ThingFO [3]. |
| | Note 3: A Test Requirement can include details of test environment requirements, which always refer to Test Context Entities. |
| **Test Result** | It is a Work Product [4] that represents both an Incident (Artifact) and an Actual Result (Outcome), which is produced by running the Perform Testing activity. |
| **Test Specification** | It is an Artifact that represents Test Checklists, Test Cases and their grouping in Test Suites. Note: Test Specification as per [9] is the "*complete documentation of the test design, test cases and test procedures for a specific test item*". |
| **Test Suite** (synonym: Test Set) | It is a Test Specification that includes a set of one or more Test Cases with common constraints on their realization. Note: Test Suite in [10] is a synonym of the Test Set term in [9]. |
| **Testable Entity** (synonym: Test Object) | A concrete object able to be tested. |
| | Note 1: Testable Entity may have none or many Test Items, which in turn are testable. |
| | Note 2: Depending on the Test Particular Situation implied by the Test Goal, Testable Entity has semantic of Developable Entity (from FRsTDO [8]) or Evaluable Entity (from NFRsTDO [6]). |
| **Testing** (synonym: Testing Process) | It is a process (Work Process [4]) that is composed of at least three interrelated Testing Activities conducted to facilitate the discovery of defects and/or the assessment of Characteristics and Attributes [6] of a Testable Entity. |
| **Testing Activity** | It is an Activity [4] that is formed by an interrelated set of sub-activities and tasks, aimed at designing, |
| | realizing or analyzing the testing endeavor for a particular Testable Entity. |
| **Testing Agent** | It is an Agent [4] –i.e., a work resource- assigned to a Testing Activity in compliance with one or more Testing Roles.<br><br>Note: A Testing Agent may use Testing Tools. |
| **Testing Automated Agent** | It is a Testing Agent, which is not a human being. |
| **Testing Design Method** | It is a Testing Method for a task included in a Design Testing activity.<br><br>Note: A Testing Design Method may use a Test Basis, while enacting the Design Testing activity. |
| **Testing Human Agent** | It is a Testing Agent, which is a human being. |
| **Testing Life Cycle** | The series of phases that a Test Project passes through from its initiation to its closure. |
| **Testing Management** | It is the set of managerial processes and activities intended to achieve the Test Goal operationalized by a Test Project. |
| **Testing Method**<br><br>(synonym: Testing Technique) | A specific and particular way to perform the specified steps for a task included in a Testing Activity.<br><br>Note 1: Testing Method is a Method as per [4].<br><br>Note 2: The specific and particular way of a Testing Method –i.e., *how* the specified steps in a testing task should be made- is represented by a design or Realization Procedure and rules. |
| **Testing Realization Method** | It is a Testing Method for a task included in a Testing Realization activity. |
| **Testing Role** | It is a Role [4] that implies a set of testing skills.<br><br>Note: Testing skills are abilities, competencies and responsibilities involved in the Testing process. |
| **Testing Strategy** | Principles, patterns, and particular test domain concepts and framework that can be specified by a set of core Testing Processes, in addition to a set of appropriated Testing Methods and Tools, as core resources, for helping to achieve the Project's Test Goal purpose. |
| **Testing Tool** | It is a Tool [4] that partially or totally accomplishes the automatic execution of a Testing Method. |

*Amount of Own or Extended Terms: 44 --- Amount of Completely Reused Terms: 4*

## Testing Component – TestTDO v1.3's Attributes or Properties

| Term | Attribute | Definition |
|------|-----------|------------|
| **Actual Result** | **value** | Numerical or categorical result. |
| **Incident** | **version** | Unique identifier, which indicates the level of evolution of the Incident artifact. |
| **Realization Procedure** | **specification** | A formal or semiformal representation of a Realization Procedure. |
| **Test Basis** | **name** | Label or name that identifies the Test Basis artifact. |
| | **version** | Unique identifier, which indicates the level of evolution of the Test Basis artifact. |
| **Test Case** | **precondition** | Any kind of constraint that must evaluate to true before the Test Case's input be used in a Perform Testing activity. |
| | | Note 1: The required state of a test item and its environment prior to test case execution [10]. |
| | | Note 2: Test case preconditions include test environment, existing data (e.g. databases), software under test, hardware, etc. [9]. |
| | **input** | Test Case inputs are the data (numbers, categories, objects' instances, etc.) and/or events (e.g., button clicks) used to exercise a Testable Entity in a Perform Testing activity. |
| | | Note: Inputs are the data information used to drive test execution [9]. |
| | **expected result** | Value and/or behavior that is expected to get after a Testable Entity is exercised using the Test Case inputs. |
| | **postcondition** | Any kind of constraint that must evaluate to true after the Test Case's input was used and the Actual Result was yielded in a Perform Testing activity. |
| **Test Checklist** | **item description** | An unambiguous textual statement describing an element of the Test Checklist. |
| **Test Conclusion Report** | **name** | Label or name that identifies the Test Conclusion Report. |
| | **version** | Unique identifier, which indicates the level of evolution of the Test Conclusion Report. |
| | **description** | An unambiguous textual statement describing the Test Conclusion Report. |

| Test Context Entity | name | Label or name that identifies the Test Context Entity. |
| --- | --- | --- |
| | description | An unambiguous textual statement describing the Test Context Entity. |
| **Test Goal** | label | Label that identifies a Test Goal uniquely. |
| | statement | An explicit declaration of the aim to be achieved.<br><br>Note: A statement is usually a written assertion in a high-level or natural language. |
| | purpose | The rationale for achieving the specified Test Goal.<br><br>Note: Examples of test purposes are: verify, validate, find defects, non-compliances, or vulnerabilities, etc. |
| | success criteria | The set of conditions by which the Test Goal will be judged as successful for stakeholders. |
| **Test Information Need** | label | Label that identifies a Test Information Need uniquely. |
| | statement | An explicit declaration of the aim to be achieved.<br><br>Note: A statement is usually a written assertion in a high-level or natural language. |
| | purpose | The rationale for achieving the specified Test Information Need goal, which basically consists on analyze to provide information. |
| **Test Particular Situation** | positive statement | An explicit declaration of a Test Particular Situation to be defined, which can refer to a static or dynamic situation.<br><br>Note 1: Regarding a particular Thing, a descriptive, positive statement refers to what it is, was, or will be, and contains no indication of approval or disapproval.<br><br>Note 2: For a Test Particular Situation, the positive statement refers to things and relationships of Testable and Test Context Objects.<br><br>Note 3: A positive statement should be based on current or subsequent empirical evidence. |
| | model specification | It represents an Artifact that specifies and models Test Particular Situations in a given language [12].<br><br>Note 1: A Test Particular Situation model has the semantic of Artifact, which is a term coming from ProcessCO [4].<br><br>Note 2: Test Particular Situations can be modeled using informal, semiformal or formal specification languages. |
| | conditions | Any kind of constraint that must be fulfilled in the Test Particular Situation. |

|  |  | Note 1: In the scenario testing approach, some conditions could be preconditions (or postconditions), which are any kind of constraint that must evaluate true before (or after) starting (or finishing) the execution of the Test Particular Situation (i.e., the scenario).<br><br>Note 2: This property has semantic of Constraint-related Assertion from ThingFO [3]. |
| **Test Project** | **name** | Label or name that identifies the Test Project. |
| **Test Requirement** | **label** | Label that identifies a Test Requirement uniquely. |
|  | **statement** | An explicit declaration of the Test Requirement to be satisfied.<br><br>Note: A statement is usually a written assertion in a high-level or natural language. |
|  | **testable entity phase** | It indicates the stage of the testable-entity life cycle in which the Testable Entity is.<br><br>Note: Examples of phases are "inception", "development", "deployment", "operation" and "maintenance". |
|  | **test level** | It represents a kind of test that delimits the scope of the Testable Entity and its context taking into account the Test Requirement statement.<br><br>Note: Examples of kinds of test levels commonly cited are "unit", "integration", "system" and "acceptance". We can also include the "document" test level. |
|  | **completion criteria** | The set of conditions by which the Test Requirement will be judged as complete for stakeholders. |
| **Test Result** | **name** | Label or name that identifies the Test Result work product. |
|  | **description** | An unambiguous textual statement describing the Test Result. |
| **Test Specification** | **name** | Label or name that identifies the Test Specification artifact. |
|  | **version** | Unique identifier, which indicates the level of evolution of the Test Specification artifact. |
| **Testable Entity** | **name** | Label or name that identifies the Testable Entity. |
|  | **description** | An unambiguous textual statement describing the Testable Entity. |
| **Testing** | **name** | Label or name that identifies the Testing work process. |
|  | **work** | Specification of what to do for achieving the objective of |

| | description | a Testing Work Process. |
|---|---|---|
| | | Note 1: The specification of what to do is a set of general actions, which implies both Testing Activities and Tasks. It represents what should be done instead of how it should be performed. |
| | | Note 2: The specification of the work description can be formal, semi-formal or informal. |
| **Testing Activity** | **name** | Label or name that identifies the Testing Activity. |
| | **work description** | Specification of what to do for achieving the objective of a Testing Activity. |
| | | Note: See note 1 and 2 in the work description attribute for the Testing term. |
| **Testing Agent** | **name** | Label or name that identifies the Testing Agent. |
| | **capabilities** | Set of abilities that the Testing Agent has. |
| **Testing Design Method** | **design procedure** | Arranged set of Testing Design Method's instructions or operations, which specifies how must be performed the Design Testing activity using the Test Basis, if any. |
| **Testing Method** | **name** | Label or name that identifies the Testing Method. |
| | **rule** | Set of principles, conditions, heuristics, axioms, etc. associated to the design procedure or Realization Procedure. |
| **Testing Role** | **name** | Label or name that identifies the Testing Role. |
| | **skills** | Set of capabilities, competencies and responsibilities of a role. |
| **Testing Strategy** | **name** | Label or name that identifies the Testing Strategy. |
| **Testing Tool** | **name** | Label or name that identifies the Testing Tool. |
| | **description** | An unambiguous textual statement describing the Testing Tool. |

*Amount of Attributes (Properties): 51*

## Testing Component - TestTDO v1.3's Non-taxonomic Relationships

| Relationship | Definition |
|---|---|
| **adopts** | A Testing Management process adopts a Testing Life Cycle. |
| **associates** | A Test Project associates one or more Testing Strategies. |
| **consumes (x6)** | An Analyze Test Results activity consumes one or more Test Result as work product. |
| | A Perform Testing activity consumes one or more Test Specification as artifact. |
| | An Analyze Test Results activity consumes one or more Test Specification as artifact. |
| | A Design Testing activity consumes Test Basis, if any. |
| | A Testing work process can consume one or more Test Particular Situation's model specifications as Artifacts. |
| | A Testing work process can consume one or more Test Requirement's specifications as Artifacts. |
| **deals with test environment** | A Test Particular Situation deals with none or many concrete Test Context Entities as a test environment. |
| **deals with test target** | A Test Particular Situation deals with one or more concrete Testable Entities as a test target. |
| **defines** | A Test Project defines one or several Test Particular Situations. |
| **helps to achieve** | A Testing Strategy gives support for achieving one or more Test Goals. |
| **implies** | A Test Goal implies one or more Test Particular Situations. |
| **influences** | A Test Context Entity influences one or several Testable Entities. |
| **involves** | A Testing Work Process involves one or more Testing Roles. In turn, a Testing Role may participate in one or more Testing Work Process. |
| **is assigned to (x4)** | A Static Testing Method is assigned to one or more Perform Static Testing activities. |
| | A Dynamic Testing Method is assigned to one or more Perform Dynamic Testing activities. |
| | A Testing Design Method is assigned to one or more Design Testing activities. |
| | A Testing Agent is assigned to one or more Testing Activities. |
| **is based on (x2)** | A Realization Procedure is based on none or several Test Specifications. |
| | A Test Requirement is based on none or several Test Basis. |
| **is derived in** | A Test Goal is derived in one or more Test Requirements. |
| **is linked to (x2)** | A Test Basis is linked to none or several Non-Functional Requirements. |
| | A Test Basis is linked to none or several Functional Requirements. |
| **is managed by** | A Test Project is managed by means of a Testing Management process. |
| **is supported by** | A Test Goal is supported by one or more Test Information Needs. |
| **operationalizes** | A Test Project operationalizes one or more Test Goals. |
| **plays** | A Testing Agent plays one or more Testing Roles. In turn, a Testing Role is played by one or more Testing Agents. |
| **produces (x5)** | A Testing Management process produces a Test Plan as artifact. |
| | A Perform Testing activity produces a Test Result as work product. |
| | An Analyze Test Results activity produces a Test Conclusion Report as artifact. |
| | A Design Testing activity produces one or more Realization Procedure. |
| | A Design Testing activity produces one or more Test Specification as artifact. |
| **refers to (x2)** | A Test Requirement in its statement always refers to a Testable Entity. |
| | A Test Requirement can refer to a one or more Test Context Entities. |
| **relies on** | An Incident detected in a Perform Testing activity relies on none or several Actual Results. |
| **requires as input (x2)** | A Testing process needs a Testable Entity as input. |
| | A Testing process needs none or many Test Context Entities as input. |
| **surrounded by** | A Testable Entity is surrounded by one or several Test Context Entities. |
| **takes into account** | An Analyze Test Results activity consumes one or more Test Information Needs as a specification. |
| **uses (x2)** | A Testing Life Cycle uses one or more Testing Strategies. |
| | A Testing Agent uses Testing Tools, if any. |
| **verifies/validates** | A Test Specification verifies/validates one or more Testable Entities. In turn, a Testable Entity is verified/validated by one or more Test Specifications. |

*Amount of non-taxonomic relationships:  43*

# Testing Component - TestTDO v1.3's Axioms

**A1** description: *For any Perform Testing activity that produces a Test Result, this result is therefore an Actual Result or an Incident, but not both at the same time.*

**A1** specification:

$$\forall prt, \forall tr : [PerformTesting(prt) \land TestResult(tr) \land produces(prt, tr)$$
$$\rightarrow ActualResult(tr) \veebar Incident(tr)]$$

**A2** description: *Any Testing process has at least three different activities, namely: Design Testing, Perform Testing and Analyze Test Results.*

**A2** specification:

$$\forall p, \exists a_1, \exists a_2, \exists a_3 : [Testing(p) \land (a_1$$
$$\neq a_2) \land (a_1 \neq a_3) \land (a_2 \neq a_3) \land DesignTesting(a_1) \land PerformTesting(a_2)$$
$$\land AnalyzeTestResults(a_3) \land partOf(a_1, p) \land partOf(a_2, p) \land partOf(a_3, p)]$$

**A3** description: *For any Perform Functional Dynamic Testing activity that consumes a Test Specification, this specification is therefore based on a Functional Requirement.*

**A3** specification:

$$\forall pfdt, \forall ts, \exists tb, \exists tdm, \exists fr, \exists dt : [PerformFunctionalDynamicTesting(pfdt) \land TestSpecification\ (ts)$$
$$\land consumes(pfdt, ts)$$
$$\rightarrow TestBasis(tb) \land TestingDesignMethod(tdm) \land FunctionalRequirement(fr)$$
$$\land DesignTesting(dt) \land isLinkedTo(tb, fr) \land consumes(dt, tb) \land isAssignedTo(tdm, dt)$$
$$\land produces(dt, ts)]$$

**A4** description: *For any Perform Non-Functional Dynamic Testing activity that consumes a Test Specification, this specification is therefore based on a Non-Functional Requirement.*

**A4** specification:

$$\forall pnfdt, \forall ts, \exists tb, \exists tdm, \exists nfr, \exists dt : [PerformNonFunctionalDynamicTesting(pnfdt)$$
$$\land TestSpecification\ (ts) \land consumes(pnfdt, ts)$$
$$\rightarrow TestBasis(tb) \land TestingDesignMethod(tdm) \land NonFunctionalRequirement(nfr)$$
$$\land DesignTesting(dt) \land isLinkedTo(tb, nfr) \land consumes(dt, tb) \land isAssignedTo(tdm, dt)$$
$$\land produces(dt, ts)]$$

**A5** description: *Any Testable Entity is an Evaluable Entity iff the Test Requirement that refers to this Thing is linked to a Non-Functional Requirement.*

**A5** specification:

$$\forall te, \exists tr, \exists tb, \exists nfr : [TestableEntity(te) \land EvaluableEntity(te)$$
$$\leftrightarrow TestRequirement(tr) \land TestBasis(tb) \land NonFunctionalRequirement(nfr)$$
$$\land refersTo(tr, te) \land isBasedOn(tr, tb) \land isLinkedTo(tb, nfr)]$$

**A6** description: *Any Testable Entity is a Developable Entity iff the Test Requirement that refers to this Thing is linked to a Functional Requirement.*

**A6** specification:

$$\forall te, \exists tr, \exists tb, \exists fr : [TestableEntity(te) \land DevelopableEntity(te)$$
$$\leftrightarrow TestRequirement(tr) \land TestBasis(tb) \land FunctionalRequirement(fr) \land refersTo(tr, te)$$
$$\land isBasedOn(tr, tb) \land isLinkedTo(tb, fr)]$$

**A7** description: *Any Test Result produced by a Perform Testing activity has at least one related Test Specification which is consumed by the same Perform Testing activity.*

**A7** specification:

$$\forall tr, \forall prt, \exists ts : [TestResult(tr) \land PerformTesting(prt) \land produces(prt, tr) \rightarrow TestSpecification(ts) \land consumes(prt, ts)]$$

**A8** description: *All Test Project operationalizes a Test Goal and associates a Testing Strategy iff this Testing Strategy helps to achieve the operationalized Test Goal.*

**A8** specification:

$$\forall tp, \forall tg, \forall ts: [TestProject(tp) \wedge TestGoal(tg) \wedge TestingStrategy(ts) \wedge operationalizes(tp, tg) \wedge associates(tp, ts) \leftrightarrow helpsToAchieve(ts, tg)]$$

**A9** description: *For all Test Requirements derived from a Test Goal, there is at least one Test Project that operationalizes this Test Goal.*

**A9** specification:

$$\forall tr, \exists tg, \exists tp: [TestRequirement(tr) \wedge TestGoal(tg) \wedge isDerivedIn(tg, tr) \\ \rightarrow TestProject(tp) \wedge operationalizes(tp, tg)]$$

**A10** description: *If a Specification-based Method is assigned to a Design Testing activity that produces a Test Specification, then always consumes a Test Basis which is used by the Specification-based Method without using the internal structure of the Testable Entity.*

**A10** specification:

$$\forall dt, \forall spbm, \forall ts, \exists tb, \exists te: [DesignTesting(dt) \wedge Specification\_basedMethod(spbm) \wedge \\ TestSpecification(ts) \wedge isAssignedTo(spbm, dt) \wedge produces(dt, ts) \rightarrow TestBasis(tb) \wedge \\ TestableEntity(te) \wedge consumes(dt, tb) \wedge \neg requiresAsInput(dt, te)]$$

**A11** description: *If a Perform Testing activity consumes a Test Case in order to produce an Actual Result, and the value of the Actual Result doesn't match with the Test Case's expected result, then the Perform Testing activity produces an Incident that relies on this Actual Result.*

**A11** specification:

$$\forall prt, \forall tc, \forall ar, \forall er, \forall val, \exists i: [PerformTesting(prt) \wedge TestCase(tc) \wedge ActualResult(ar) \\ \wedge ExpectedResult(er) \wedge Value(val) \wedge consumes(prt, tc) \wedge partOf(er, tc) \\ \wedge produces(prt, ar) \wedge partOf(val, ar) \wedge (er \neq val) \\ \rightarrow Incident(i) \wedge produces(prt, i) \wedge reliesOn(i, ar)]$$

**A12** description: *If a Structure-based Method is assigned to a Design Testing activity that produces a Test Specification, then always requires as input the internal structure of the Testable Entity that is used by the Structure-based Method.*

**A12** specification:

$$\forall dt, \forall stbm, \forall ts, \exists te: [DesignTesting(dt) \wedge Structure\_basedMethod(stbm) \wedge TestSpecification(ts) \wedge \\ isAssignedTo(stbm, dt) \wedge produces(dt, ts) \rightarrow TestableEntity(te) \wedge requiresAsInput(dt, te)]$$

**A13** description: *If a Testing process requires as input a Testable Entity, then some of its Testing Activities require and use it as input as well.*

**A13** specification:

$$\forall p, \exists te, \exists ta: [Testing(p) \wedge TestableEntity(te) \wedge requiresAsInput(p, te) \\ \rightarrow TestingActivity(ta) \wedge partOf(ta, p) \wedge requiresAsInput(ta, te)]$$

**A14** description: *If a Testing process requires as input a Test Context Entity, then some of its Testing Activities require and use it as input as well.*

**A14** specification:

$$\forall p, \forall tce, \exists ta: [Testing(p) \land TestContextEntity(tce) \land requiresAsInput(p, tce) \\ \rightarrow TestingActivity(ta) \land partOf(ta, p) \land requiresAsInput(ta, tce)]$$

**A15** description: *If a Testing process consumes a Test Requirement's specification, then some of its Testing Activities consume it as well.*

**A15** specification:

$$\forall p, \forall trs, \exists ta: [Testing(p) \land SpecificationOfTestRequirement(trs) \land consumes(p, trs) \\ \rightarrow TestingActivity(ta) \land partOf(ta, p) \land consumes(p, trs)]$$

**A16** description: *If a Testing process consumes a Test Particular Situation's specification, then some of its Testing Activities consume it as well.*

**A16** specification:

$$\forall p, \forall tps, \exists ta: [Testing(p) \land SpecificationOfTestParticularSituation(tps) \land consumes(p, tps) \\ \rightarrow \exists ta: TestingActivity(ta) \land partOf(ta, p) \land consumes(p, tps)]$$

**A17** description: *If a Testing process involves a Testing Role, then some of its Testing Activities involve it as well.*

**A17** specification:

$$\forall p, \forall tr, \exists ta: [Testing(p) \land TestingRole(tr) \land involves(p, tr) \\ \rightarrow TestingActivity(ta) \land partOf(ta, p) \land involves(p, tr)]$$

*Amount of axioms: 17*

# Appendix A: Updates from TestTDO v1.2 to TestTDO v1.3

Note that the TestTDO v1.2 ontology is documented in [2]. The main changes are:

- We changed the name of the Testing Realization activity, which now is called "Perform Testing". However, we maintained Testing Realization as a synonym.
- We changed the name of the Testing Design activity, which now is called "Design Testing". However, we maintained Testing Design as a synonym.
- We changed the name of the Testing Analysis activity, which now is called "Analyze Test Results". However, we maintained Testing Analysis as a synonym.
- We changed the name of the Static Testing activity, which now is called "Perform Static Testing". However, we maintained Static Testing as a synonym.
- We changed the name of the Dynamic Testing activity, which now is called "Perform Dynamic Testing". However, we maintained Dynamic Testing as a synonym.
- We changed the name of the Functional Dynamic Testing activity, which now is called "Perform Functional Dynamic Testing". However, we maintained Functional Dynamic Testing as a synonym.
- We changed the name of the Non-Functional Dynamic Testing activity, which now is called "Perform Non-Functional Dynamic Testing". However, we maintained Non-Functional Dynamic Testing as a synonym.
- We updated the definitions of concepts, properties, non-taxonomic relationships and axioms according to these newly used names in the Testing Activities.
- We included now the concepts of NFRsTDO and FRsTDO into the TestTDO packet as completely reused terms.
- We added in the stereotypes the namespaces corresponding to each ontology of the FCD-OntoArch, which semantically enriches each term.
- We added the property "conditions" in Test Particular Situation.
- Since in SituationCO the Particular Situation pattern was updated, we applied the changes according to this new update in TestTDO in the terms Test Particular Situation, Testable Entity and Test Context Entity. This involved adding two new non-taxonomic relationships, namely: "deals with test environment" and "deals with test target".
- We updated the references. Now we quoted current related works.
- Figure 1 was updated accordingly.